\documentclass{hotnets17}

\usepackage{times}  
\usepackage{hyperref}

\usepackage{epsfig}
\usepackage[TABBOTCAP]{subfigure}
\usepackage{tabularx}
\usepackage{graphicx}
\usepackage{color}
\usepackage{xspace}
\usepackage{thumbpdf}
\usepackage{listings}
\usepackage{verbatim}
\usepackage{booktabs}
\usepackage{colortbl}
\usepackage{caption}

\hypersetup{pdfstartview=FitH,pdfpagelayout=SinglePage}

\usepackage{booktabs}
\usepackage{authblk}
\usepackage{etoolbox}

\definecolor{grey}{rgb}{0.5,0.5,0.5}
\begin{document}



\title{Micro Congestion Control: Every Flow Deserves a Second Chance}

\author[*]{Kefan Chen}
\author[*]{Danfeng Shan}
\author[*]{Xiaohui Luo}
\author[*]{Tong Zhang}
\author[+]{Yajun Yang}
\author[+]{Ya Zhao}
\author[*]{Fengyuan Ren}

\affil[*]{Department of Computer Science, Tsinghua University\authorcr ckf16@mails.tsinghua.edu.cn}
\affil[+]{Tencent Technology Shenzhen Company}
\maketitle

\begin{abstract}

Today, considerable Internet traffic is sent from the datacenter and heads for users. 
The characteristics of connections served by servers in datacenters are usually diverse and varied over time, with continuous upgrades in network infrastructure and user devices. As a result, a specific  congestion control algorithm hardly accommodates the heterogeneity and performs well in various scenarios. In this work, we present \textbf{M}icro \textbf{C}ongestion \textbf{C}ontrol (\textbf{MCC}) --- a novel framework for Internet congestion control. With MCC, diverse algorithms can be assigned purposely to connections in one server to adapt to heterogeneity, and different algorithms can be chosen in each connection's life cycle to keep pace with the dynamic of network. We design and implement MCC in Linux, and the experiments validate that MCC is capable of smoothly switching among various candidate algorithms on the fly to achieve potential performance gain in the real world. Meanwhile, the overheads introduced by MCC are moderate and acceptable.

\end{abstract}

\section{Introduction}
Due to the Internet's large-scale resource-sharing nature and continuous technology evolution in network infrastructure and user devices, the principle and techniques of congestion control have been unceasingly studied for decades  \cite{vanjacobson}. Meanwhile, to support massive and various online applications where users' quality of experience is severely affected by network transmission performance, datacenter to user is witnessed as the prevalent architecture. Among all networked applications in datacenters, large content transfers (e.g., high-resolution videos) dominate the network traffic, and congestion control scheme directly determines the performance of such traffic. 

In datacenter-to-user architecture, one server usually faces clients whose network path characteristics are diverse and dynamic \cite{yan1622pantheon}. To accommodate this heterogeneity, plenty of efforts have been devoted. For instance, dozens of classical congestion control algorithms are proposed and implemented in the Linux kernel, but server operators usually configure a unified algorithm for all connections in a server\cite{naseer2017configtron}. Another thread of efforts, which also targets to accommodate the heterogeneity, evolves towards revolutionizing the methodology of congestion control \cite{remy, pcc2015, pcc2018}. They follow the creed that the handcrafted approach fail to cope with heterogeneity and react ineffectively to the dynamic reality, resulting in less ideal performance. To make the congestion control algorithm adaptive to various network environment, Remy \cite{remy}
utilizes the machine-generated congestion control rules to replace manually designed algorithms, but the rules are mined from the offline data of given network condition, which can not completely summarize the dynamic of network. PCC  \cite{pcc2015, pcc2018} is a performance-oriented rate control architecture that adjusts the sending rate to maximize its utility value defined by combined performance metrics.

Traditional handcrafted algorithms may fail to attain the ideal performance when they are exposed to the dynamic and diversity of Internet. Unlike other attempts, we open another door inspired from the following facts. (1) The large-scale online applications residing in datacenters, especially for those with long-lived connections, pose requirements for congestion control policy to account for heterogeneity, meanwhile, it also provides the opportunity to optimize congestion control performance by mining volume network feedback data.
(2) One specific congestion control algorithm, even the state of the art, cannot excel in diverse scenarios, as described in \textsection{2.1}. 
Therefore, we do not intend to design an omnipotent algorithm to excel in all scenarios. Instead, we take a less radical reformism approach: with the fact that a set of algorithms outperform others in some specific  scenarios, we attribute the rigidity of applying the same default algorithm for all connections to the lack of fresh knowledge about each connection's network environment, and we pin the performance degradation on the mismatch of algorithm adoption. We seek to  bridge the gap by identifying the connection's network characteristics and properly selecting the suited congestion control algorithm.

In this paper, we introduce \textbf{M}icro \textbf{C}ongestion \textbf{C}ontrol (\textbf{MCC}) --- a framework for Internet congestion control, whose main function is to select the proper algorithm for each connection. It has two "micro" features. (1) Spatial: diverse congestion control algorithms can be assigned to different connections in one server. (2) Temporal: different algorithms can be applied in different phases of each connection's life cycle.
We implement MCC in Linux and the preliminary evaluations show that MCC can provide the following benefits: (1) MCC ensures smooth and live transition among various algorithms; (2) MCC performs better than the unified congestion control algorithm; (3) MCC introduces modest overheads.

\textbf{Roadmap.} In \textsection{2}, we state our motivation and challenges to realize MCC, then describe how to design and implement MCC in \textsection{3} and \textsection{4}, respectively. \textsection{5} presents a case study about selecting suited algorithm. Preliminary evaluations are presented in \textsection{6}. \textsection{7} and \textsection{8} discusses and summarizes our work.

\begin{figure}[!t]
	\centering
	\setlength{\belowcaptionskip}{-0.5cm}	
	\includegraphics[width=7cm]{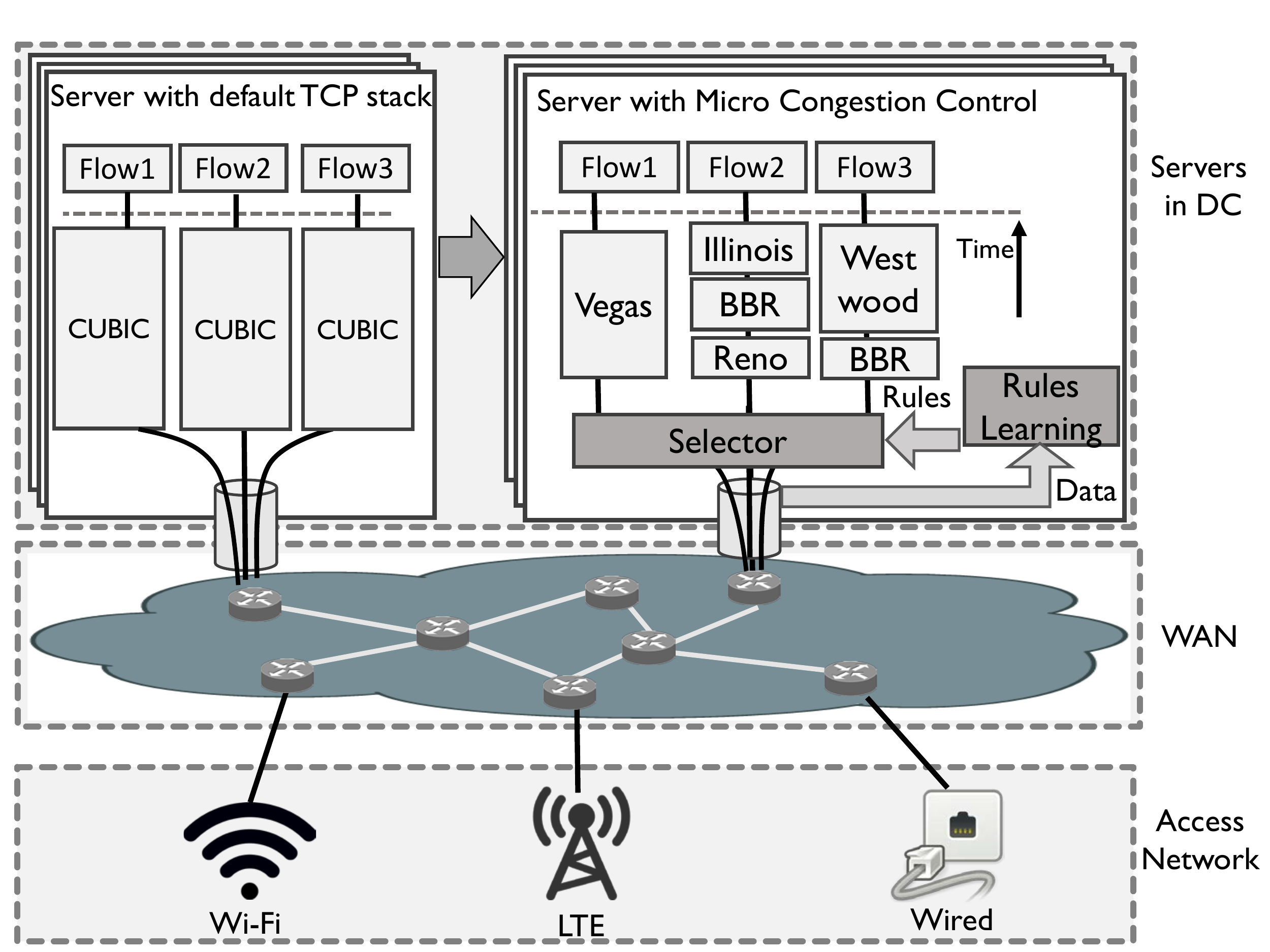}
	\captionsetup{font={scriptsize},labelfont=bf}
	\caption{\textbf{Basic idea of MCC}}
	\label{basicidea}
	\vspace{-0.1cm}
\end{figure}

\vspace{-0.1cm}
\section{Motivation and Challenge}
\vspace{-0.2cm}
\subsection{Motivation}
The servers in datacenters face clients whose connections have diverse and dynamic network characteristics. The heterogeneity is largely attributed to the diversity of access network, carrier network and access time. Furthermore, especially for long-lived connections, the characteristics are likely time-varying, which is caused by allround factors, such as competing flows' joining/departing, link failover and route change, etc. The dynamic can be exploited to improve performance, for example, by detecting the kinds of its competing flows' congestion control algorithm (loss-based like CUBIC \cite{ha2008cubic} or delay-based like Vegas \cite{brakmo1995tcp} ), then switching among modes with different degree of aggressiveness, Copa \cite{arun2018copa} exhibits the combined advantages of loss-based and delay-based algorithms, i.e., high throughput and low queuing delay.

One specific congestion control algorithm cannot always excel in diverse scenarios. 
Existing investigations demonstrate that performance depends on network environments. The findings in \cite{yan1622pantheon} tell that the performance superiority of different algorithms can vary with  network path and running time. According to results in \cite{pcc2018}, PCC outperforms most of the tested algorithms but is inferior to Sprout \cite{winstein2013stochastic} in LTE environments. Besides, "More can be less". The experimental study in \cite{sivaraman2014experimental} reveals that the RemyCC trained with TCP-awareness performs better than the RemyCC without TCP-awareness when TCP cross traffic is present, but performs worse when TCP cross traffic is absent. Therefore, adding more functionalities to a single algorithm can have side-effects when the targeted scenario is not present.

However, nowadays servers usually apply a unified congestion control algorithm for all connections without considering their respective properties\cite{naseer2017configtron}. We argue that the traditional handcrafted algorithms based on some premises may fail to attain ideal performance when they are exposed to dynamic network environment, moreover the unified congestion control algorithm is too rigid to adapt to network complexity and heterogeneity, but we do not question the effectiveness when they are against their targeted problem. For instance, TCP Westwood \cite{westwood} is specialized for wireless network,  whose performance is guaranteed by theoretic analysis. The handcrafted rationale of TCP Westwood, however, is hard for machine to learn by exploration from scratch in one single connection's life cycle. 

Therefore, we do not attempt to design an omnipotent algorithm to excel in most scenarios. Instead, we attribute the rigidity of applying the unified algorithm  for all connections to the lack of fresh knowledge about each connection's network environment, and we pin the performance degradation on the mismatch of algorithm adoption. We seek to  bridge the gap by identifying the connection's  network characteristics and selecting the suited congestion control algorithm.

To achieve this, as shown in the right part of Figure \ref{basicidea}, our basic idea contains the following steps:
(1) collecting run-time network feedback data for each connection;
(2) identifying the classes of connections; 
(3) selecting the appropriate congestion control algorithms based on the classes and rules acquired from mining history data.

Specifically, MCC is a framework which requires no receiver modification, it provides the following functions:
(1) collecting raw run-time network data from the network protocol stack; 
(2) defining an efficient interface to expose collected data to the analysis procedure which identifies each connection's class and selects a suited algorithm in a real-time manner with the given rules; 
(3) ensuring switching smoothly among various algorithms. 
In this work, we focus on designing and implementing the underlying system which enables applying the suited congestion control algorithm to each connection, the rules of mapping data to specific class and corresponding algorithm will be studied sufficiently in the future.

\vspace{-0.2cm}
\subsection{Challenges to Get MCC off the Ground}
To realize MCC in practice, the following challenges need to be addressed properly.

\textbf{Live migration between algorithms}. To apply multiple algorithms in different phases of one connection, we need to migrate the connection states as switching algorithm. The smooth algorithm switching can be nontrivial. To avoid performance degrading drastically, the initial state of the new algorithm should be chosen carefully. Besides, since different congestion control algorithms maintain individualized variables, it is difficult for the new algorithm to maintain performance if necessary variables are not updated by the previous algorithm. For instance, the observed maximum bandwidth is used to compute sending rate in BBR \cite{cardwell2016bbr}, but not all algorithms maintain this variable, such as Vegas and CUBIC.

\textbf{Overheads.}
Characterizing connections online and deciding suitable algorithms requires analyzing feedback data in a real-time manner, because existing protocol implementation usually maintains several reduced variables only, such as smooth RTT(Round-Trip Time), which are just a profile of raw data, but the evolution trace of the connection is needed to characterize it and to make classifications. During collecting and analyzing the original data(e.g., RTT, loss rate, bandwidth), extra overheads are introduced, including huge-amount data extraction and high-frequency data analysis, which may harm the performance of servers.

\vspace{-0.1cm}
\section{Design}
In this section, we present the details of designing MCC as well as how to address the above challenges. 

\begin{figure}[!t]
	\setlength{\belowcaptionskip}{-0.2cm}	
	\begin{tabular}{cc}
		\centering
		\small
		\begin{minipage}[htbp]{0.24\textwidth}
			\centering
			\includegraphics[width=3.8cm]{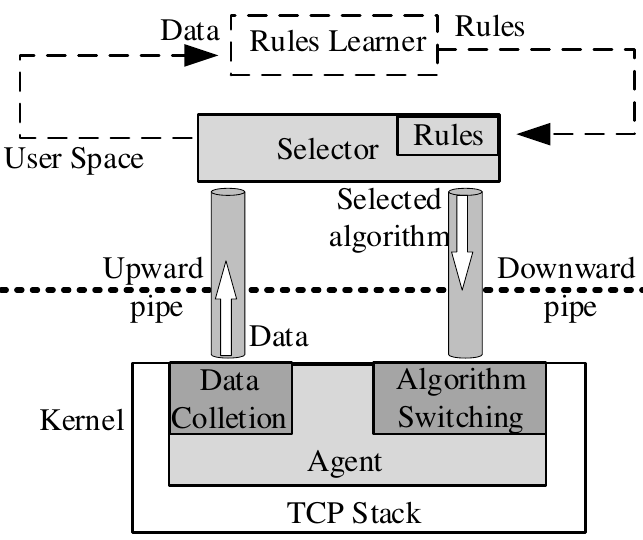}
			\captionsetup{font={scriptsize},labelfont=bf}
			\caption{\textbf{MCC Architectural Overview}}
			\label{mccarch}
		\end{minipage}
		\begin{minipage}[htbp]{0.24\textwidth}
			\centering
			\includegraphics[width=4.2cm]{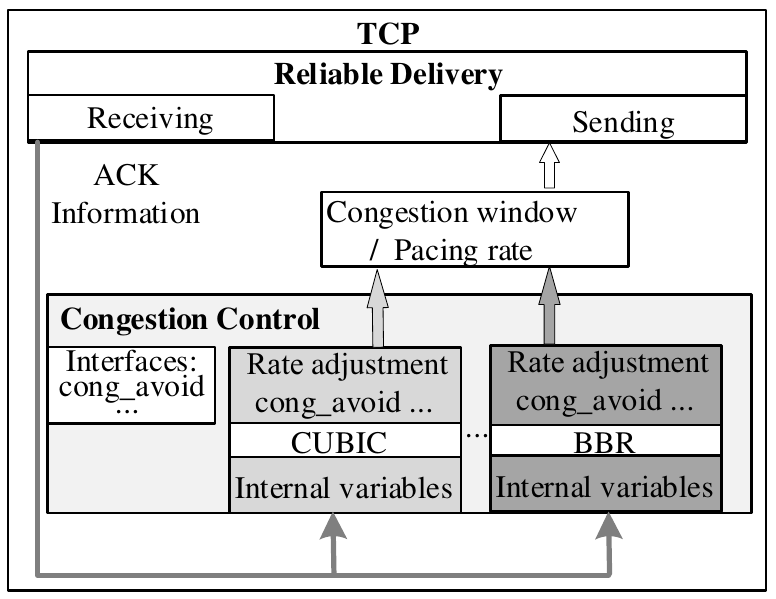}
			\captionsetup{font={scriptsize},labelfont=bf}
			\caption{\textbf{Modularized TCP}}
			\label{tcparch}
		\end{minipage}
	\end{tabular}
	\vspace{-0.4cm}
\end{figure}

Figure \ref{mccarch} shows a high-level schema of MCC, including two key modules: \emph{Selector} and \emph{Agent}. \emph{Selector} resides in user space and identifies connections' features and maps the features to corresponding congestion control algorithms according to the given rules.  \emph{Agent} is responsible for coordinating information exchange between \emph{Selector} and kernel stack. More specifically, \emph{Agent} orchestrates information exchange in two directions. The upward pipe transmits data collected by \emph{Agent} from kernel stack and exposes them to \emph{Selector}, the downward pipe conveys notifications generated by \emph{Selector}, then \emph{Agent} parses the notification and launches the algorithm switching in kernel stack.  

\emph{Selector} is put in user space since characterizing connections is computing-intensive which involves float point operation. \emph{Rules Learner} plays the role of setting rules to \emph{Selector} by mining the interested data. Its detailed designing is beyond the scope of this work, we will further investigate its concrete algorithm and implementation in future work.
\subsection{Design Components}
\vspace{-0.2cm}
\subsubsection{Agent}

\emph{Agent} mainly provides two functions: data collection and algorithm switching.

\textbf{Data collection}. 
Data-collection procedure is invoked when an ACK(Acknowledgment) is received or the retransmission timer fires. Firstly, \emph{Agent} extracts the information(e.g., current RTT, loss rate, bandwidth) and then writes them to the upward pipe where the data is read by the \emph{Selector} later.

\textbf{Algorithm switching}. 
This is the crucial functional component as well as the distinguished feature in comparison with conventional congestion control scheme. We describe it thoroughly from two aspects. (1) How to replace congestion control algorithm online? (2) How to ensure the smoothness of the transition between algorithms?

\begin{table*}[htbp]
	\setlength{\belowcaptionskip}{-0.5cm}
	\captionsetup{font={scriptsize},labelfont=bf}
	\caption{\textbf{Observed variables of classical congestion control algorithms}}
	\label{tabmeasurement}
	\scriptsize
	\centering
	\renewcommand\arraystretch{1.25}
	\begin{tabular}{l|l}
		\toprule
		\textbf{Algorithm} & \textbf{Typical observed variables and their update frequency} \\
		\midrule
		CUBIC & dealy\_min (updated when packets are ACKed); last\_max\_cwnd (updated when retransmission timer expired)\\
		BBR & min\_rtt (updated when ACK arrives); max\_bw (updated when ACK arrives)\\
		Vegas & base\_rtt (updated when ACK arrives); count\_rtt, min\_rtt (updated when packets are ACKed)\\
		Westwood & bandwidth\_estimation (updated when current sending round ends); cumulated\_acked (updated when ACK arrives)\\
		Illinois\cite{illinois} & sum\_rtt, count\_rtt (updated when packets are ACKed); 
		max\_rtt, base\_rtt (updated when packets \\
		& are ACKed); alpha, beta (updated when current sending round ends)\\
		\bottomrule
	\end{tabular}
\vspace{-0.5cm}
\end{table*}

\emph{How to replace congestion control algorithm online?} Algorithm replacement is done by replacing the reference to the algorithm\footnote{In Linux kernel, the reference to congestion control algorithm is in the form of function pointer, replacement is done by changing the function pointer to the new algorithm.}, which can be realized under the support of modularized TCP implementation, as depicted in Figure \ref{tcparch}, where congestion control algorithm is implemented as pluggable module. Specifically, the implementation architecture of modularized TCP satisfies two requirements\cite{arianfar2012tcp,cheng2016making} . (1) The congestion control module only computes the sending rate and does not involve in other TCP functions. (2) Algorithms in the congestion control module share a common congestion state machine (e.g., congestion avoidance, loss recovery) and implement corresponding defined interface of rate adjustment. Taking slow start state for example, all algorithms should implement their own rate control policy for slow start so that the congestion control module can call the interface when a connection enters slow start.
In this way, the congestion control module only interacts with the common interface of algorithms and can invoke the related rate adjustment interface in any state. Therefore, algorithm switching can be done by replacing the reference to the algorithm.

\emph{How to ensure the smoothness of the transition between algorithms?} 
The state migration for algorithm switching is essentially translating the variables of previous algorithm to the variables of new algorithm, and two kinds of variables are related to smoothness: the sending rate variable (in the form of congestion window or pacing rate) and observed variables (e.g.,  the minimum observed RTT), because the sending rate directly determines performance, and observed variables are used to adjust sending rate (e.g., BBR computes the sending rate by the maximum observed bandwidth and the minimum observed RTT). We take two measures to decide the initial value of the sending rate and observed variables of the new algorithm. (1) Inheriting the previous algorithm's sending rate to avoid drastic performance degradation. We set the initial congestion window for the new algorithm by inheriting the previous algorithm's evolutional value. As for the algorithms employing the pacing mechanism, such as BBR, we set the initial pacing rate to the value of congestion window divided by recent sampled RTT. And the transition from pacing-rate based algorithm to window-based algorithm is symmetric. (2) Initializing the new algorithm's observed variables to their default value. The reason is that the observed variables are secondary to sending rate variables in metric of performance, and the update frequency of observed variables is high enough to compensate the information loss caused by initializing them to default values. To confirm this point, some typical observed variables of Linux kernel congestion control algorithm are listed in Table \ref{tabmeasurement}. Most of them are updated per-ACK, and multiple ACKs are received in one RTT, which can generate adequate samples to update observed variables. Besides, initializing observed variables requires no transplant work for algorithms to fit in our framework, because the initialization is done by calling their original initialization function, otherwise, we need to find the transition relation between any two algorithms since observed variables of different algorithms are usually dissimilar. To sum up, the previous algorithm's sending rate is inherited and the observed variables are reset for the new algorithm to ensure the smoothness, and the evaluation result suggests the feasibility of this method in \textsection{6.1}. 


\vspace{-0.2cm}
\subsubsection{Selector}
The functions of \emph{Selector} contains: (1) reading collected data from upward pipe; (2) analyzing data with the given rules to select the algorithm for each connection; (3) sending notifications through the downward pipe to inform \emph{Agent} to switch algorithm. 

Specifically, \emph{Selector} maintains  per-connection states in its memory to store the necessary information for algorithm selection. Each time it reads collected data from upward pipe, it updates the per-connection states. After the updates, If both the per-connection states and given rules indicate another algorithm is preferable, \emph{Selector} informs \emph{Agent} to switch algorithm by sending notifications through the downward pipe.

\vspace{-0.2cm}
\subsubsection{Pipes}
The pipes transmit the data between \emph{Agent} and \emph{Selector}. Specifically, the upward pipe transmits data collected by \emph{Agent} from the TCP stack and exposes them to \emph{Selector}, the downward pipe transmits the algorithm switching notifications. It is crucial for the pipes to minimizes data exchange overheads(see details in \textsection{4.1}).

\vspace{-0.2cm}
\section{Implementation}
We implement MCC in the Linux kernel 4.14.29 and the associated user-level library. We firstly present the optimization techniques for both pipes and \emph{Selector}, then introduce the implementation details.

\subsection{Components Optimization}
\textbf{Pipes optimization.} Firstly, we choose the ring buffer as the data structure of both upward and downward pipes. \emph{Agent} is the data producer and \emph{Selector} is the consumer for upward pipe, and the downward pipe is symmetric. Then we minimize two kinds of overheads: memory allocation/release and data access. Memory pools are pre-allocated for both upward and downward pipe respectively to avoid frequent memory allocation and release. Since the rate of writing data into upward pipe is the same as ACK arriving rate. The data access overheads are decreased by memory mapping. Taking upward pipe for instance, for the data is collected in kernel and analyzed by \emph{Selector} in user space, the memory mapping avoids the overheads of user/kernel mode switch and system calls for reading and writing data.

\textbf{Selector optimization.}
We employ two techniques to optimize \emph{Selector}. (1) \emph{Batch processing.} To amortize the overheads of synchronization between \emph{Agent} and \emph{Selector}, \emph{Selector} reads a batch of data from the upward pipe, the downward-pipe case is symmetric. (2) \emph{Per-core data structure}. One \emph{Selector}, one \emph{Agent} and two pipes are localized for each core. We choose the per-core data structure as the basic thread model for two considerations: (1) ensuring the selection decision is consistent, namely a connection's data should be processed by one core only. Otherwise, if multiple cores analyze different parts of one connection's data, they may have disagreements in algorithm selection due to the limitations of partial information extracted from a fragment of data; (2) with per-core data structure, the synchronization between \emph{Selector} and \emph{Agent} can be described as the single-producer-single-consumer model, which supports lock-free implementation.

\subsection{MCC Kernel Implementation}
MCC contains two parts in kernel: an external kernel module and modifications to the TCP stack.

\vspace{-0.2cm}
\subsubsection{Kernel Module}
\emph{Selector} interacts with \emph{Agent} through a special device file: /dev/mcc. \emph{Selector} calls \texttt{open} system call to open the file to create a suite of pipes, 	which is also accessible from \emph{Agent}. A handle is returned to \emph{Selector} for manipulating pipes. Besides, the pre-allocated memory pool is allocated by the driver of /dev/mcc in the \texttt{open} system call. In addition, the driver implements the \texttt{ioctl} system call to realize the synchronization between \emph{Selector} and \emph{Agent}, and the memory mapping is achieved by implementing the \texttt{mmap} interface of the device. Pipes are maintained as lock-free ring structures which are pre-allocated and memory-mapped.  Batch processing is implemented by \emph{Selector} polling the pipes, that is, \emph{Agent} continuously writes collected data to the shared area, where \emph{Selector} periodically reads a batch of data by timer-triggering. The timer interval naturally becomes a prominent factor that directly influences the batch size and data freshness. Currently, we adopt an empirical approach of setting the timer interval around two milliseconds, because data is carried by ACK and several ACKs are arrived in one RTT for each connection, and RTT is typically millisecond-level in Internet, and two milliseconds is far larger than the CPU cycles consumed by one timer waking up operation.

\vspace{-0.2cm}
\subsubsection{Kernel TCP Stack Modification}
The minor modifications are made to Linux kernel TCP stack to support \emph{Agent}'s functions, including data collection and algorithm switching.

\textbf{Data collection}. We instrument the ACK processing procedure to generate raw data, and extract per-ACK information, e.g., RTT, loss rate, delivery rate, ECE ( ECN-Echo ) mark, etc. More information can be elicited if required.

\textbf{Algorithm switching}. 
Dozens of congestion control algorithms are implemented in Linux kernel in the form of loadable modules. Changing the congestion control algorithms of connections can be realized by replacing function pointers. The switching starts by \emph{Selector} writing algorithm decision of a group of connections in the downward pipe and calling \texttt{ioctl}, which notifies \emph{Agent} to set algorithm switching flags for the related connections' TCP control blocks. The flag will directly trigger an algorithm replacement when kernel congestion control component is invoked. The state migration for the replacement is in accordance with \textsection{3.1.1}.

\subsection{MCC User Interface}
MCC user library is essentially a simple wrapper of the kernel module, which includes the \texttt{mcc\_open} and\\ \texttt{mcc\_close} to open/close the device file, and \texttt{ioctl} to synchronize \emph{Agent} and \emph{Selector}. The pseudocode in \textsection{5} shows an example of using the interface.

\section{A Case Study about MCC Rules}
we present a case study about classifying WiFi connections from wired connections to: (1) provide an intuitive example about MCC rules; (2) demonstrate the workflow and usage of MCC interface; (3) validate that potential performance can be gained by MCC (\textsection{6.2}). The classification accuracy  and more abundant rules will be further explored in the future work.

\vspace{-0.1cm}
\captionsetup{font={scriptsize},labelfont=bf}
\begin{lstlisting}[language={c},caption={\textbf{Pseudocode for WiFi classification}},captionpos=b,basicstyle=\scriptsize]
/*Selector*/
/*get a handle to manipulate MCC*/
handle = mcc_open();
/*main loop of classifcation*/
while(true) {
/*update upgoing pipe for new data*/
 ioctl(handle->fd, UPSYNC,NULL);
/*whether Agent has writen data in the upward pipe*/
 while(!pipe_empty(handle->up_pipe)) {
  data = get_data(handle->up_pipe)
/*get the state of the flow who own the data*/
  flow_state = hash_find(data.flow_id);

/*the rule to classify WiFi in the seciton 5*/
  if(flow_state.rtt_cnt < N && 
   flow_state.newcc == 0) {
   update_state(flow_state, data);
   if(flow_state.rtt_cov > CTH && 
    flow_state.rtt_range > RTH) {
    flow_state.newcc = westwood;
/*write algorithm switching message to downward pipe*/
    put_data(handle->down_pipe, flow_state));
   }
  }
 }

/*signal Agent to switch algorithm, if any*/
 ioctl(handle->fd, DOWNSYNC, NULL);
/*sleep to achieve batch processing*/
 sleep(2ms);
}
\end{lstlisting}
\vspace{-0.2cm}

We summarize the rules to differentiate WiFi connections from wired connections by analyzing their trace, and find that the jitter of RTT is drastic in WiFi (illustrated in Figure \ref{wifi}) and we quantize it by the coefficient of variation (COV, standard deviation divided by mean) and normalized range ( (maximum - minimum)/minimum ) of sampled RTTs. The rule is described as following: if the COV and the normalized range of sampled RTTs for a connection exceed certain thresholds (CTH and RTH in the pseudocode) in the first N sampled RTTs, we speculate the connection is through WiFi and set TCP Westwood for this connection.

\section{Preliminary Evaluation}
We answer three questions in this section. Whether smooth and online algorithm switching is feasible? Whether potential performance gain can be achieved by MCC? Whether the overheads are moderate?

\subsection{Algorithm Switching}

\begin{figure*}[!t]
		\begin{tabular}{ccc}
		\centering
		\small
	\parbox{0.33\linewidth}{
		\includegraphics[width=\linewidth]{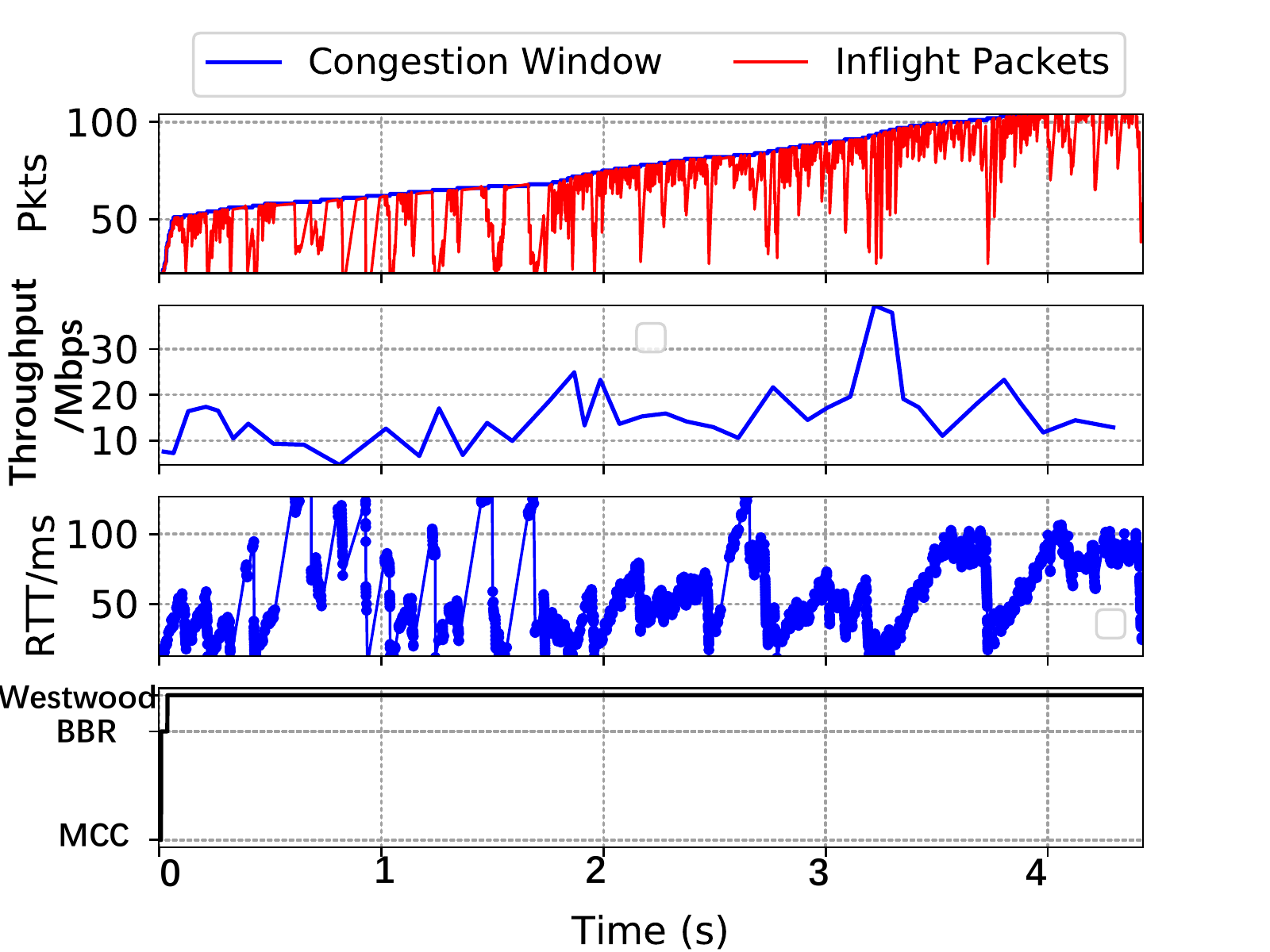}
		\captionsetup{font={scriptsize},labelfont=bf, justification=centering}
		\caption{\textbf{MCC in the WiFi environment}}
		\label{wifi}		
	}
	\parbox{0.33\linewidth}{
		\includegraphics[width=\linewidth]{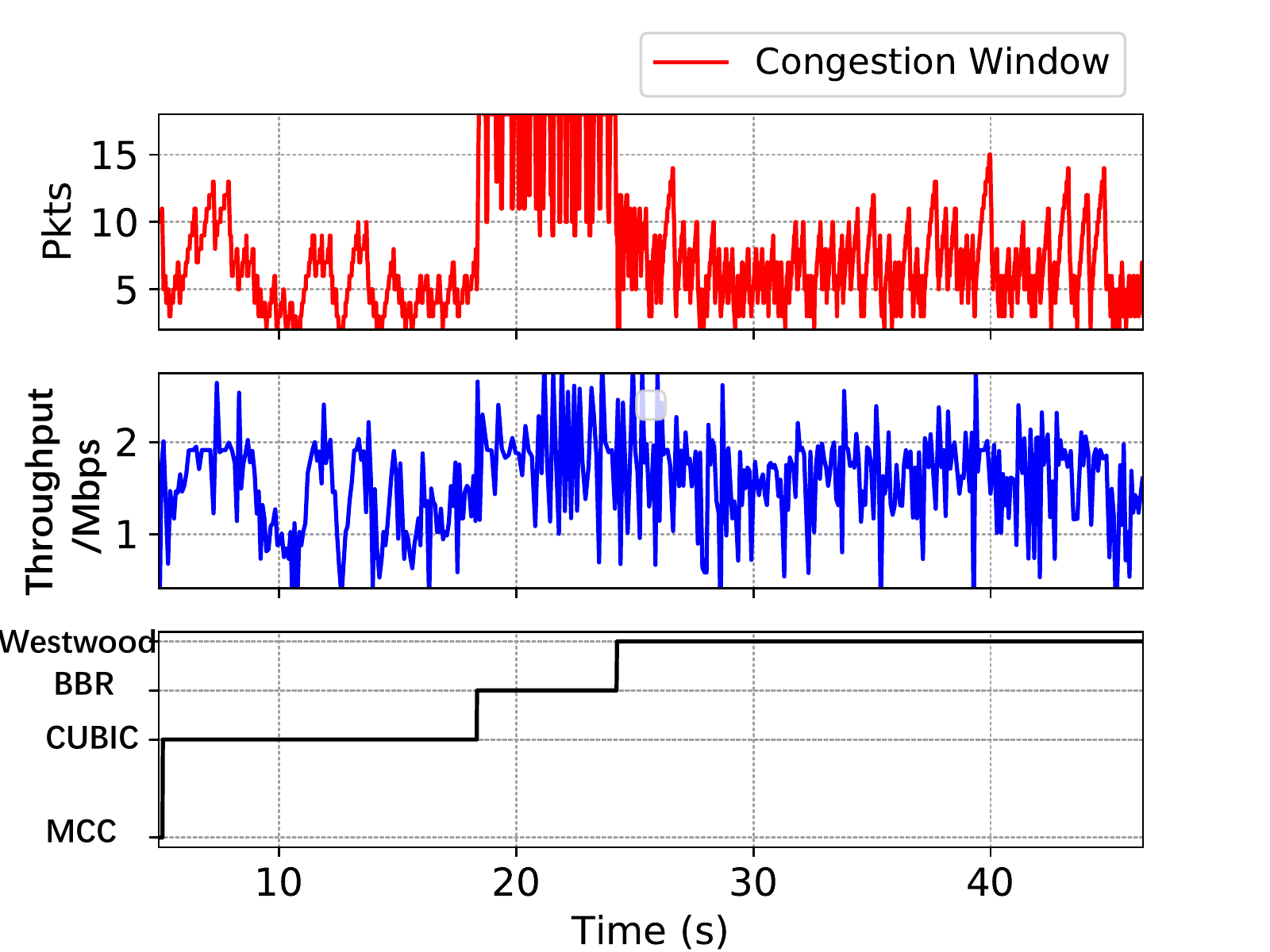}
		\captionsetup{font={scriptsize},labelfont=bf, justification=centering}
		\caption{\textbf{Algorithm switching}}
		\label{migration}		
	}

	\parbox{0.33\linewidth}{
		\includegraphics[width=\linewidth]{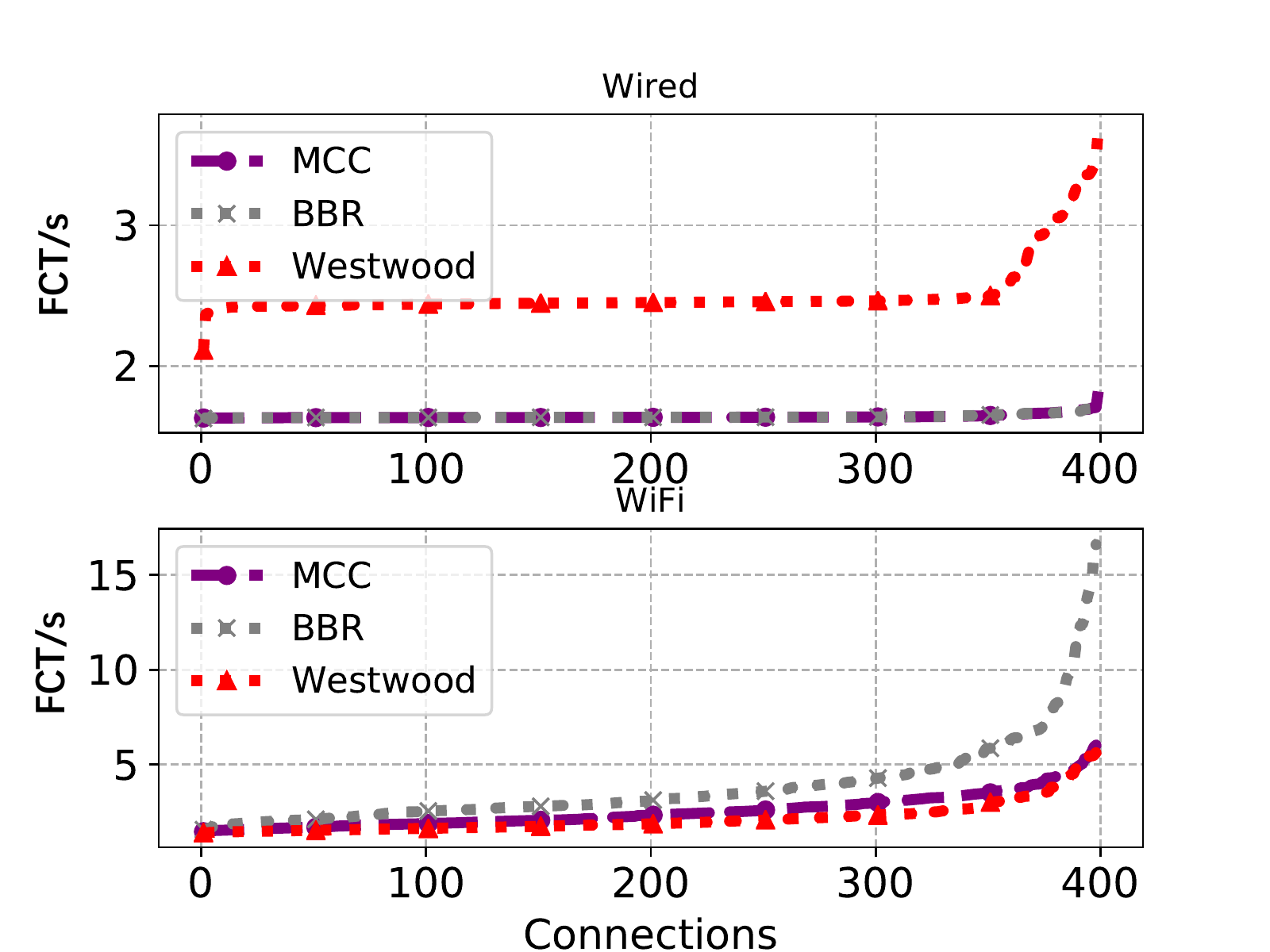}
		\captionsetup{font={scriptsize},labelfont=bf, justification=centering}
		\caption{\textbf{Sorted flow completion time}}
		\label{perfgain}	
	}	
		
	\end{tabular}
	\vspace{-0.5cm}
\end{figure*}

To observe the dynamic behavior of MCC during algorithm switching, we firstly utilize the Linux Traffic Control (TC) \cite{linuxtc} to regulate the link characteristics to: 2 Mbps bandwidth, 30ms RTT and 4\% loss ratio. Driven by MCC,the congestion control algorithm is switched from CUBIC to BBR, then to Westwood, which covers the dominated algorithm types, i.e., rate-based/window-based, delay-based/loss-based. TCP probe \cite{tcpprobe} is then used to trace the following metrics: sender congestion window and throughput. The results are shown in Figure \ref{migration}. Obviously, the smooth and online algorithm switching can be conducted by MCC.

\vspace{-0.1cm}
\subsection{Performance Gain}
We show that MCC averagely outperforms BBR by 31.0\% and Westwood by 11.4\% in the heterogeneous network by creating two real network scenarios: clients with WiFi access and clients with wired access. The server resides in the University lab running three Nginx \cite{nginx} web servers with MCC, BBR and Westwood, respectively, two client machines send requests to all three web servers to download 8MB files. One client machine accesses a public WiFi, the other is a virtual host in the cloud platform. The default algorithm of MCC is BBR, and the rule developed in \textsection{5} is used to differentiate WiFi connections from wired connections. BBR, Westwood and MCC are tested at the same time to reduce bias, and the related performance metrics are collected over a day. Figure \ref{perfgain} shows the sorted flow completion time of MCC, BBR and Westwood. In the wired environment,  MCC and BRR surpass Westwood. In wireless environment, MCC performs better than BBR and is close to Westwood. The difference between Westwood and MCC is due to the classification algorithm only analyze the first N sampled RTT, the RTT jitter of wireless connection can be mild at first but severe later, in this case, BBR is selected and perform less ideal due to it misunderstands the fluctuation of RTT as the signal of queueing. The average flow completion time of MCC, BBR and Westwood in two scenarios are 2.10 s, 2.75 s, 2.34 s, respectively.

\subsection{Overheads}
\vspace{-0.2cm}
\begin{table}[htbp]
	\captionsetup{font={scriptsize},labelfont=bf}
	\caption{\textbf{CPU cost percentages comparison of MCC and netlink}}
	\label{cpu_cost}
	\scriptsize
	\centering
	\begin{tabular}{c|cc|cc}
		\toprule
		\  &\multicolumn{2}{c|}{\textbf{User space / \%}} &\multicolumn{2}{c|}{\textbf{Kernel / \%}} \\	
		\midrule
		MCC &\emph{Selector} 0.54 & Nginx 29.24 & \emph{Agent} 0.08 & \texttt{tcp\_ack} 0.38 \\
		netlink &nl\_user 11.83  & Nginx 21.13  & nl\_kernel 0.54 & \texttt{tcp\_ack} 0.71\\
		\bottomrule
	\end{tabular}
	\vspace{-0.5cm}
\end{table}

In this section, we test the overheads of MCC in high-load server. To validate our system optimizations, we compare 
MCC with netlink socket \cite{netlink}, which is designed and commonly used for transferring miscellaneous networking information between the kernel and user space processes. Its API complies with socket, which is inefficient in the scenarios of handling high-speed short messages (per-ACK data in our context) due to factors such as kernel/user mode switching and data copy \cite{mtcp}.

We set up one server machine and one client machine, which are both equipped with a 12-core CPU (Intel Xeon E52620 @ 2.4GHz), 64GB RAM and one dual-port Intel 82599 10G NIC, respectively.  The 10G port of 82599 NIC in the server is directly connected to client machines' NIC port. We stress the CPU by combining the multi-queue of 10G NIC into a single queue, and bind the single queue to one specific core in the server. Then MCC is applied to Nginx \cite{nginx} and ensure MCC, Nginx, kernel network stack all run in the same core. To identify the overheads caused by MCC itself, \emph{Selector} only pre-processes data by computing average RTT, loss rate, and throughput, computation of characterizing connection is not introduced. The client machine firstly regulates the RTT to 20 ms, then 250 concurrent persistent connection are launched in the client using an HTTP benchmark tool (wrk) \cite{wrk} to download 1MB files, which nearly saturates the 10G NIC port. We replace MCC's pipes and related enhancements with netlink socket as the comparison experiment.

We then profile the CPU cycles cost proportions by Linux performance profiling tool \cite{linuxperf} to evaluate the overheads of MCC and netlink socket. As listed in Table \ref{cpu_cost}, 
MCC contains two parts, \emph{Selector} in user space and \emph{Agent} in kernel. To make the CPU cost proportion value more concrete, we compare MCC's overheads with Nginx and \texttt{tcp\_ack} procedure. (Nginx runs in user space as the application while \texttt{tcp\_ack} is the original kernel procedure we add \emph{Agent} in). The netlink socket also includes two parts, nl\_user, which reads data from kernel, and nl\_kernel, which is instrumented into \texttt{tcp\_ack} to extract data. The result shows that the overheads introduced by MCC are modest while netlink's cost is nonnegligible.


\section{Discussion}
\vspace{-0.1cm}
\textbf{Dimensions of MCC.} MCC exploits the heterogeneity  ("spatial", among connections) and variability ("temporal", within a connection), we preliminarily discuss their relationships. (1) "spatial" focuses on the intrinsic properties of the flow, e.g., access network characteristics, while "temporal" catches the run-time pathological behaviors, e.g., traffic policing \cite{policing}. (2) The "spatial" classification is triggered in the start of a flow (the algorithm in \textsection{5}), and the "temporal" classification is triggered when the specified symptoms arise, thus, "spatial" is more general.

\textbf{Rules learning.} In future work, we plan to generate the rules by combining two methodologies: white-box method (WB) and black-box method (BB), where the former refers to rules summarized by human domain knowledge (the algorithm in \textsection{5}), the latter refers to rules acquired by machine learning. The two methods are complementary in the way that WB is specialized and with low overheads, while BB is generalized and resource-consuming. 

\textbf{Data-driven congestion control.} Remy, PCC and Copa agree that the space of congestion control signals and actions is too complicated for handcrafted algorithms and choose to search the optimum solution by numerical computation, their control unit granularity is more subtle (e.g., sending rate), while MCC's control unit (algorithm) is more stable and understandable because of their design rationale (e.g., Westwood for wireless network). 

\textbf{Fairness.} The fairness is largely determined by the objective implied in the rules and the selected algorithm, we will investigate it further in future work.

\vspace{-0.2cm}
\section{Conclusion}
\vspace{-0.1cm}
In this paper, we argue that, instead of applying a unified congestion control algorithm for connections with heterogeneous network, preferable algorithm should be selected according to the characteristics of each connection. We design and implement the prototype of MCC to support our argument, it aims to improve performance by exploiting the heterogeneity and variability of network. Preliminary evaluations confirm the feasibility of MCC. Future work will focus on designing and implementing \emph{Rules Learner} and deploying MCC in the datacenter to attain real performance gain. 


\bibliographystyle{abbrv} 
\begin{small}
\bibliography{hotnets17}
\end{small}

\end{document}